\documentclass[osajnl,twocolumn]{revtex4-1}
\usepackage[psamsfonts]{amssymb}
\usepackage{amsmath}
\usepackage{color}
\usepackage{graphics}
\begin{document}
\title{The critical role of the energy spectrum in determining the nonlinear-optical response of a quantum system}
\author{Shoresh Shafei and Mark G. Kuzyk}
\affiliation{Department of Physics and Astronomy, Washington State University, Pullman, Washington 99164-2814\\
Corresponding author: shafei@wsu.edu}

\begin{abstract}
Studies aimed at understanding the global properties of the hyperpolarizabilities have focused on identifying universal properties when the hyperpolarizabilities are at the fundamental limit.  These studies have taken two complimentary approaches: (1) Monte Carlo techniques that statistically probe the full parameter space of the Schrodinger Equation using the sum rules as a constraint; and, (2) numerical optimization studies of the first and second hyperpolarizability where models of the scalar and vector potentials are parameterized and the optimized parameters determined, from which universal properties are investigated.  Here, we employ an  energy spectrum constraint on the Monte Carlo method to bridge the divide between these two approaches.  The results suggest an explanation for the origin of the factor of 20-30 gap between the best molecules and the fundamental limits and establishes the basis for the three-level ansatz.

\end{abstract}

\maketitle

\section{Introduction}

The fundamental limit of the off-resonant electronic hyperpolarizability and second hyperpolarizability are calculated using the constraints on the energies and matrix elements of position operator (which we loosely call them transition moments throughout this text) imposed by the Schrodinger Equation.\cite{kuzyk00.01,kuzyk00.02,kuzyk01.01}  It is most convenient to quantify these constraints in the form of the Thomas-Kuhn sum rules, and then using the assumption that when first and second hyperpolarizabilities, $\beta$ and $\gamma$ respectively, are optimized only three states (including the ground state) contribute - called the three-level ansatz. The upper limit of $\beta$ is given by,\cite{kuzyk00.01,kuzyk00.02,kuzyk01.01}
\begin{equation}\label{betamax}
\beta_{max} = 3^{1/4} \left(\frac{e \hbar}{\sqrt{m}}\right)^3 \left(\frac{N^{3/2}}{E_{10}^{7/2}}\right) ,
\end{equation}
where $-e$ is charge of the electron, $\hbar$ is Planck's constant, $N$ is the number of electrons of the quantum system and $E_{10}$ is the energy difference between the first excited state and ground state. The second hyperpolarizability follows along the same lines as for the hyperpolarizability,\cite{kuzyk00.01,kuzyk00.02,kuzyk03.01,kuzyk03.02} and yields,
\begin{equation}\label{gammamax}
-\left( \frac {e \hbar} {\sqrt{m}} \right)^4 \frac {N^2} {E_{10}^5} \leq \gamma_0 \leq 4 \left( \frac {e \hbar} {\sqrt{m}} \right)^4  \frac {N^2} {E_{10}^5} \equiv \gamma_{max}.
\end{equation}

A comparison of the largest experimentally measured hyperpolarizability with the fundamental limit reveals a large gap between the two. Measurements have never crossed the limit and are typically well below it.\cite{kuzyk03.01,kuzyk03.02,Tripa04.01,Tripa07.01}  Until about 2007, the hyperpolarizabilities of all molecules ever measured mysteriously fell a factor of about 30 below the limit.  As such, the limits were seen to provide an absolute metric of the nonlinear-optical response.  This idea was used, for example, by Slepkov and coworkers to understand the surprising nonlinear optical properties of polyene oligomers;\cite{slepk04.01} and by May and coworkers to show the promise of small molecules.\cite{May05.01,May07.01} Chen applied the limits to study the nonlinear response of nano-engineered polymers.\cite{wang04.01}  More recently, a new class of twisted molecules were reported by Kang and coworkers that appear to have exceptionally large hyperpolarizabilities.\cite{zhou08.01,kuzyk09.01,Kang05.01,Kang07.01}

A more important consequence of the limits is the fact that they provide a method for determining how the nonlinear-optical response scales with the quantum size of the system.  It is straightforward to show that if the scalar potentials and energies are re-scaled by a factor $\mu^{-2}$ and the vector potential and positions are all rescaled by a factor $\mu$, the shape of the wavefunction remains the same aside from it being compressed by the scaling factor $\mu$.\cite{kuzyk10.01} This transformation leaves the intrinsic hyperpolarizabilities, defined by,
\begin{equation}\label{intrinsic 1st & 2nd hyperpolarizability}
\beta_{int} = \frac {\beta} {\beta_{max}} \hspace{1em} \mbox{and} \hspace{1em} \gamma_{int} = \frac {\gamma} {\gamma_{max}}
\end{equation}
invariant. As such, the intrinsic hyperpolarizabilities can be used to compare molecules of very different shapes and sizes.

The concept of scale invariance suggests that optimization of the hyperpolarizability is a two-step process.  First, a successful paradigm is identified that optimizes the intrinsic hyperpolarizability. Subsequently, a quantum system within that paradigm can be made larger using simple scaling.  As an example, Roberts and coworkers have shown that triphenylamine-cored alkynylruthenium dendrimers have a nonlinear-optical response - as characterized by two-photon absorption (TPA) cross section per electron - that is about an order of magnitude larger than the non-triphenylamine-cored versions.\cite{rober09.01} However, when the results are properly scaled according to the intrinsic two-photon cross-section,\cite{kuzyk03.03,perez05.01} all of the triphenylamine-cored dendrimers had the same intrinsic value that was found to be two orders of magnitude larger than the non-triphenylamine-cored versions.\cite{perez11.01} Thus, the triphenylamine-cored dendrimer is the new molecular paradigm with a larger intrinsic TPA cross-section than other systems; and, dendrimers with triphenylamine-cores can then be made larger by adding additional dendrimer generations (i.e. additional branches) to increase the absolute TPA cross-section.

Scale invariance has defined two approaches for building an understanding of the nature of a quantum system when the hyperpolarizability is near the limit, both of which seek to optimize the intrinsic quantities.  The first are studies that use a parametrization of the scalar and vector potentials to determine the nature of the potentials that lead to an optimized nonlinear response.\cite{kuzyk06.02}  This approach has been used to suggest a new paradigm for making molecules\cite{zhou06.01,zhou07.02} that has lead to the synthesis and characterization of a record intrinsic hyperpolarizability.\cite{perez07.01,perez09.01} Such studies also show that a quantum system near the fundamental limit shares certain universal properties.\cite{watkins09.01} Wang and coworkers have used related optimization techniques to theoretically build larger molecules from smaller molecular building blocks as the basic units.\cite{wangm06.01}

A second approach is using Monte Carlo simulations that statistically probe the full space of allowed solutions to the Schrodinger Equation by using the Thomas-Kuhn sum rules as a constraint.\cite{kuzyk08.01,shafei10.01}  In contrast to the potential function approach, which yields an optimized intrinsic hyperpolarizability of 0.78, the Monte Carlo approach yields a distribution of hyperpolarizabilities that approaches arbitrarily close to the fundamental limit.  The Monte Carlo approach is the most general, but, gives no information of how to design a quantum system with a large hyperpolarizability.  Our present work seeks to bridge the divide between the Monte Carlo approach and real systems by classifying materials according to their energy spectrum.

In contrast to studying specific system to gain an understanding of the origins of the nonlinear optical response,\cite{hefli98.01,hefli98.02,wu89.01} our work seeks to identify broad principles with the aim of applying these results to finding new paradigms of the nonlinear optical response.\cite{kuzyk09.01}

\section{Approach}

The Thomas-Kuhn sum rules, which are a direct consequence of the Schrodinger Equation, are given by,
\begin{equation}\label{SR}
\sum_{n=0}^{\infty} \left[E_n - \frac{1}{2}(E_m + E_p)\right] x_{mn}x_{np} = \frac{\hbar^2 N}{2 m}\delta_{mp} ,
\end{equation}
where $\delta_{ij}$ is the Kronecker delta, $E_m$ is the energy of state $m$ and $x_{nm}$ is the $(n,m)$ position matrix element.  Equation \ref{SR} holds for multi-electron Hamiltonians with $N$ electrons with spin, externally applied electromagnetic fields, and electron correlations.\cite{kuzyk10.01}  The relativistic sum rules, which are generalizations of Equation \ref{SR}, \cite{Goldman82.01, Leung86.01, cohen04.01} are not required for most molecular systems. Sum rules have also been used to study the dispersion of the nonlinear response;\cite{keina08.01} but, our present interest is in the off-resonant regime.

For simplicity, we use the dimensionless sum rules
\begin{equation}\label{sumrules}
\sum_{n=0}^{\infty} \left[e_n - \frac{1}{2}(e_m + e_p)\right] \xi_{mn}\xi_{np} = \delta_{mp},
\end{equation}
where $e_i = E_{i0}/E_{10}$ (the ground state is labeled by $0$) and the normalized transition moments are given by
\begin{equation}\label{transitionMoments}
\xi_{ij} = \frac{x_{ij}}{|x_{01}^{max}|}
\end{equation}
where
\begin{equation}\label{x01max}
|x_{01}^{max}|^2 =\frac{\hbar^2 N}{ 2mE_{10}}.
\end{equation}
Equation \ref{x01max} defines the upper bound of the transition moment from the ground state, $x_{01}$, by use of Equation \ref{SR} in light of the fact that $E_{10} < E_{i0}$ for $i>1$.

The numerical procedure we apply here is similar to the previous studies of the hyperpolarizability,\cite{kuzyk08.01} and the second hyperpolarizability,\cite{shafei10.01} with the exception that the energy levels are not chosen randomly.  Instead, we set the energies to be of a predefined functional form,
\begin{equation}\label{EnergyFunction}
E_s = f(s).
\end{equation}
For a particular function $f(s)$, the Monte Carlo simulations span a restricted set of solutions of the Schrodinger Equation that gives a statistical sampling of the solutions that have a particular energy spacing.  This approach of fixing the energy-level spacing is motivated by the observation that detuning in real systems from optimal energy spacing is responsible for the factor of 30 gap between the best molecules and the fundamental limit.\cite{Tripa04.01,perez07.02,Tripa07.01}

The energy values in our model must be chosen for states $j>i$ such that $E_j > E_i$. Thus, for energy classes such as $E_j = -j^{-1}$, we shift all energies so that this constraint holds. As an example,
\begin{equation}\label{energyShift}
E_{j} = -\frac{1}{j} \rightarrow -\frac{1}{j+1} + 1
\end{equation}
Thus, the ground state is labeled by $j=0$ and has zero energy.  Since $e_{i} = E_{i0}/E_{10}$, then $e_{0}=0$ and $e_{1}=1$.

The transition moments are assigned, as in our previous work \cite{kuzyk08.01,shafei10.01}, so that the energies and transition moments are consistent with sum rules. Since we use the dipole-free expression for $\beta$ \cite{kuzyk05.02} and $\gamma$ \cite{perez01.08} we need only use diagonal sum rules ($m=p$) in Equation \ref{sumrules} to get all of the required transition moments.  Starting with $(m,p)=(0,0)$, we get
\begin{equation}\label{m=p=0}
e_{10}|\xi_{10}|^2 + e_{20}|\xi_{20}|^2 + e_{30}|\xi_{30}|^2 + \ldots + e_{n0}|\xi_{n0}|^2 = 1.
\end{equation}
$e_{10} = e_1 - e_0 = 1$, so $\xi_{01}$ is randomly assigned from the interval $ -1 < \xi_{01} <  1$. Subsequently, $\xi_{02}$ is obtained from
\begin{equation}\label{sumrule00-1}
\sum_{n=2} e_{n0}\xi_{n0}^2 = 1 - e_{10}\xi_{10}^2,
\end{equation}
using
\begin{equation}\label{ineq1}
\xi_{20}^2 \leq \left( 1 - e_1 \xi_{10}^2 \right)/e_{20}.
\end{equation}
A random number $-1 \leq r \leq 1$ is used to get $\xi_{20}$ from Equation \ref{ineq1},
\begin{equation}\label{rule2}
\xi_{20} = r\sqrt{\left( 1 - e_1 \xi_{10}^2\right)/e_{20}}.
\end{equation}
For a system with s-states, the procedure is repeated for all other transition moments, $\xi_{i0}$, except for $\xi_{s-1,0}$, which is directly determined from the last remaining diagonal sum rule.

The next set of transition moments are calculated using the higher-order sum rules in sequence with $p = 1,2,3,\ldots$ in Equation \ref{sumrules}. Since $\xi_{ij}$ is assumed to be real for all states $i$ and $j$,
\begin{equation}
\xi_{ij} = \xi_{ji}.
\end{equation}
The energy and transition moment values are then used to calculate the intrinsic first and second hyperpolarizabilities.  This procedure is iterated repeatedly to generate a distribution of first and second hyperpolarizability values.  The process is then repeated for a variety of energy level functions.

Using the dipole-free sum over state expression \cite{kuzyk05.02}, $\beta$ in the off-resonant regime is given by
\begin{equation}\label{betaDF}
\beta = -3 e^3 {\sum_{n,m}^{\infty}}' x_{0n}x_{nm}x_{m0} \left(\frac{1}{E_{n0} E_{m0}} - \frac{2E_{m0}-E_{n0}}{E_{n0}^2 E_{m0}}\right)
\end{equation}
where a prime indicates the ground state is excluded from the summation. The second hyperpolarizability, $\gamma$ is given by
\begin{eqnarray} \label{gammaDF}
\gamma &=& \frac{1}{8} \left( 2 {\sum_{n}^{\infty}}' {\sum_{m\neq n}^{\infty}}' {\sum_{l \neq n}^{\infty}}' \left\{ \frac{(2E_{m0}-E_{n0})(2E_{l0}-E_{n0})}{E_{n0}^{5}} \right. \right. \nonumber \\
&& \left. - \frac{(2E_{l0}-E_{n0})}{E_{m0}E_{n0}^{3}} \right\} x_{0m}x_{mn}x_{nl}x_{l0} \nonumber \\
&& + 2 {\sum_{n}^{\infty}}' {\sum_{m \neq n}^{\infty}}' {\sum_{l \neq m}^{\infty}}' \left\{\frac{1}{E_{l0}E_{m0}E_{n0}} \right.\nonumber \\
&& \left. -\frac{(2E_{l0}-E_{m0})}{E_{m0}^{3}E_{n0}}\right\} x_{0l}x_{lm}x_{mn}x_{n0} \nonumber \\
&&\left. -  {\sum_{m}^{\infty}}'{\sum_{n}^{\infty}}' \left\{\frac{1}{E_{m0}^{2}E_{n0}}  + \frac{1}{E_{n0}^{2}E_{m0}} \right\}x_{0m}^{2}x_{0n}^{2} \right)
\end{eqnarray}

\section{Results and Discussion}

We apply Monte Carlo simulations to eight classes of energies, $-j^{-3}$, $-j^{-2}$, $-j^{-1}$, $j^{1/2}$, $j$, $j^2$, $e^j$ and $e^{-j}$. The nature of their distributions can be used to determine what properties are relevant for optimizing the nonlinear response.

\subsection{The Three-Level Ansatz}

The three-level ansatz states that when the nonlinear response of a quantum system is at the fundamental limit, only three states contribute.\cite{kuzyk10.01}  It can be rigourously shown that the linear response is optimized for a two-level system.  However, for a nonlinear response, an analysis of the two-level model shows that it violates the sum rules.  The three-level ansatz was used in calculating the fundamental limits based on the argument that an optimized system concentrates the oscillator strength in the least number of states.  Since the three-level model is the lowest-order approximation that obeys the sum rules, it was used without proof.\cite{kuzyk00.01}

\begin{figure}
  \includegraphics{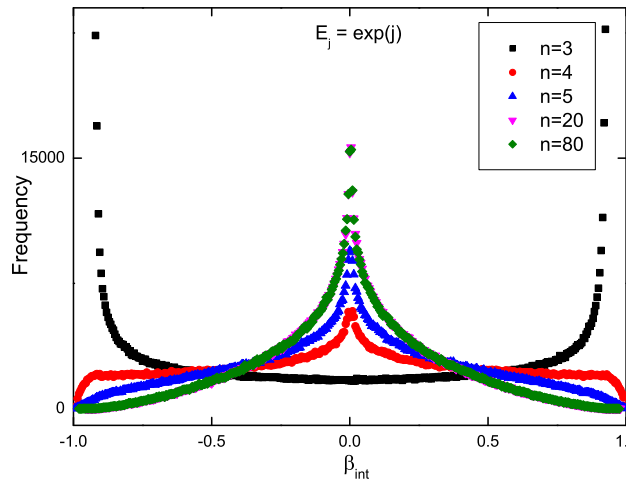}
  \caption{Distribution of $\beta_{int}$ for energy function $E_j \propto \exp(j)$.}\label{Beta,e^j}
\end{figure}
\begin{figure}
  \includegraphics{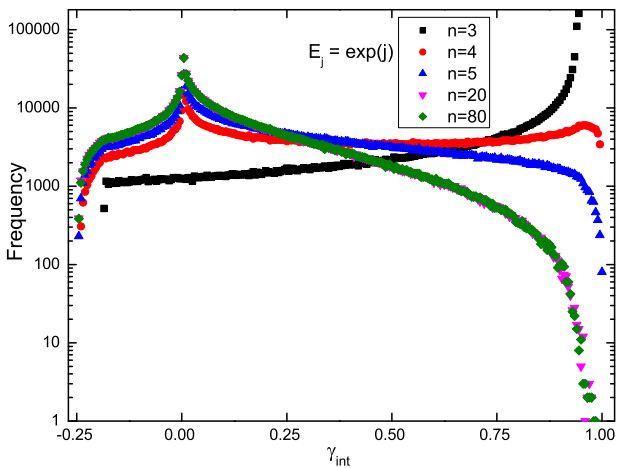}
  \caption{Distribution of $\gamma_{int}$ for energy function $E_j\propto \exp(j)$.}\label{Gamma,e^j}
\end{figure}
While the three-level ansatz has not been proven analytically, it appears to always hold in a large number of simulations.  We find the same results here.  Figure \ref{Beta,e^j} shows the distribution of $\beta$ for an exponential energy function $E_j = e^j$.  Shown are runs for 3 to 80 states where the distributions for 20 and 80 states overlap.  Thus, adding more states will not change the character of the distribution.  The three-level model shows a distribution that peaks at the limits while all of the other distributions peak at $\beta_{int} = 0$ in a cycloid-like function, a result that we observed in previous studies.  Interestingly, all these distributions tail off dramatically at $\beta_{int} = 1$.  These results are clearly consistent with the three-level ansatz.

Figure \ref{Gamma,e^j} shows the distribution of $\gamma_{int}$, also for an exponential energy function $E_j = e^j$.  Note that the second hyperpolarizability has a different positive and negative limit.  As in the case of $\beta$, the distribution of the three-level model peaks at the fundamental limit, where the second hyperpolarizability is positive.  Again, all of the other distributions peak at $\gamma_{int} = 0$.  Thus, the three-level ansatz also holds for the second hyperpolarizability.

The shape of the distribution changes dramatically near $\beta_{int} = 1$ and $\gamma_{int} = 1$ as a function of the number of states.  For $n=3$, the function is strongly upward sloping even on the log plot, showing that statistically, a three-level system with exponential energy spacing is most likely to be found near the fundamental limit.  In the four-level model, the distribution is flat over most of the domain, suggesting that all values of the intrinsic hyperpolarizabilities are equally likely, but is strongly sloped downward near the limit.  As the number of states is increased, the distribution becomes more strongly peaked near zero first and second hyperpolarizabilities.  This behavior shows how a many-level system will almost always have a smaller intrinsic first and second hyperpolarizabilities than a three-level system.

\subsection{Hyperpolarizability}

Analytical calculations of the fundamental limits using the three-level ansatz suggest that the ideal system is one in which the ratio $E_{10} / E_{20} \rightarrow 0$, i.e. when the second excited state energy is much larger than the first.  This type of energy-level spacing is not typically found in real quantum systems.  For a particle in an infinite square well, the energy scales as $E_j = j^2$ and for a harmonic oscillator it scales as $E_j = j$.  For a single electron and point nucleus, on the other hand, the energy scales as $E_j = j^{-2}$.  We define an energy class by the functional form of the energy level spacing.  Thus, the harmonic oscillator falls in the energy class $E_j = j$ and one electron atoms in $E_j = j^{-2}$.

Calculation of the fundamental limit of the hyperpolarizability leads to \cite{kuzyk00.01,kuzyk05.02},
\begin{equation}\label{betafG}
\beta = 6 \sqrt{\frac {2} {3}} e^3 \frac {\left| x_{10}^{MAX} \right|^3} {E_{10}^2} G(X) f(E) = \beta_0 G(X) f(E) ,
\end{equation}
where $\beta_0 = \beta_{max}$,
\begin{equation}\label{DEFf(E)}
f(E) = (1-E)^{3/2} \left( E^2 + \frac {3} {2} E + 1 \right),
\end{equation}
and
\begin{equation}\label{defG(X)}
G(X) = \sqrt[4]{3} X \sqrt{\frac {3} {2} \left( 1 - X^4\right)},
\end{equation}
with $X = x_{10} / x_{10}^{MAX}$ and $E = E_{10} / E_{20}$.  At the limit, $G(X) = 1$ and $f(E) = 1$.

Each energy class corresponds to a specific value of $E$ and therefore a specific value of $f(E)$.  In the case of the three-level model, each random assignment of transition moments simply identifies one value of $X$. For example, for the exponential energy class, $E = 0.27$ and $f(E) = 0.922$.  The distribution of intrinsic hyperpolarizabilities asymptotically approaches $\beta_{int} = 0.922$ for those cases where $G(X)$ = 1.  For the three-level model, the largest value of the intrinsic hyperpolarizability for a particular energy class is thus $\beta_{int} = \beta/\beta_{max} =f(E)$. Thus, $f(E)$ can be interpreted as the largest intrinsic value that is allowed by the three-level ansatz.

\begin{figure}
  \includegraphics{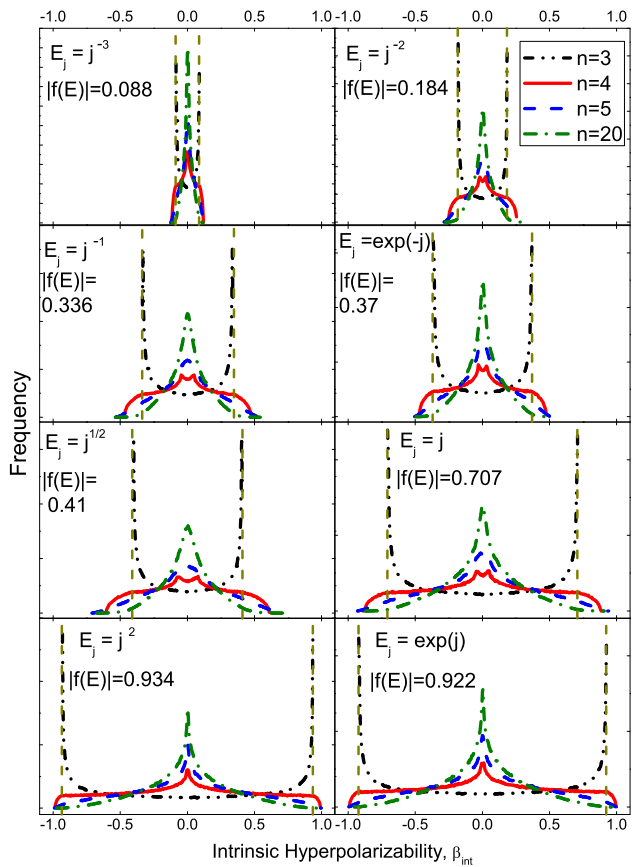}
  \caption{Distribution of $\beta_{int}$ as a function of the number of states for several different energy functions. The vertical dashed lines represent $f(E)$ for that energy function.}\label{fig:beta}
\end{figure}

Figure \ref{fig:beta} summarizes the Monte Carlo simulations of several energy classes. The dashed vertical lines represent $\beta_{int} = \pm f(E)$.  The three-level model is thus observed to behave as predicted -- the ceiling in the intrinsic hyperpolarizability gets larger as $E$ increases.

In general, an $n$-level model will have $n-2$ parameters.\cite{zhou07.02} Thus, one would expect that with more parameters, it may be possible to find combinations that yield a larger intrinsic hyperpolarizability than the three-level model.  Interestingly, the three-level model yields a distribution that peaks at $\beta_{int} = \pm f(E)$.  Thus, if the energy spacing is optimized, the distribution peaks at $\beta_{int} = \pm 1$.  In contrast, when more states are included, the distribution peaks at $\beta_{int} = 0$ and falls off sharply at $\beta_{int} = \pm f(E)$.  Thus, while the energy function $f(E)$ was specifically defined for the three-level model, it appears to define the cutoff for any number of levels for any quantum system that obeys the sum rules.

There are practical implications of these observations.  First, when a quantum system is designed to concentrate oscillator strength in only two excited states, from a statistical perspective, it is most likely that the resulting hyperpolarizability will be near the limit.  If the oscillator strength is shared amongst many states, then the most likely hyperpolarizability is zero.  Since molecules typically have many states with large transition moments, it is not surprising that the hyperpolarizabilities fall far short of the fundamental limit.  Note that beyond 20 states, all the resulting distributions appear identical.

When the energy function is suboptimal, the intrinsic hyperpolarizability can exceed the value obtained from a three-level model, that is, $\beta_{int} > f(E)$.  This suggests that it may be possible to make a quantum system with a large intrinsic hyperpolarizability in which many states contribute.  It is therefore interesting to investigate the proprieties of the outliers in the distribution to determine if this may lead to a new paradigm for making molecules with a large hyperpolarizability.

\begin{table}\caption{Summary of the properties of the hyperpolarizability for various energy functions when the hyperpolarizability is small or near the limit.}\label{tab:betaSummary}
\begin{tabular}{c c c c c}
  \hline
  Function & $\beta_{int}$ & Levels & Dominant & Maximum \\
     &  &  & States & $\beta_{nm}^{int}$\\
  \hline
$\exp(j)$ & -0.981 & 3 & 1,3 & $\beta_{12}^{int} = -0.78$ \\
$j$ & -0.90  & 4 & 1,8,9 & $\beta_{18}^{int} = -0.56$ \\
$j^{-2}$ & 0.284 & 9 & 1,4,5,6,7,8,11,12 & $\beta_{56}^{int} = -0.310$ \\
\hline
\end{tabular}
\end{table}

Table \ref{tab:betaSummary} summarizes the properties of an outlier for the energy class $\exp(j)$, in which the hyperpolarizability exceeds the three-level model ($\beta_{int} = -0.981$ and $|f(E)| = 0.922$), the energy class $j$ with $|f(E)|=0.707$ and class of energy $j^{-2}$, a case were the intrinsic hyperpolarizability is much less than the for a three-level system.  We use $\beta_{nm}$, a single term in the double sum in Equation \ref{betaDF}, to represent the fractional contribution to the hyperpolarizability of the two states $n$ and $m$.  The total hyperpolarizability is then given by,
\begin{equation}\label{betanm}
\beta = \sum_{n,m} \beta_{nm} .
\end{equation}

\begin{figure}
  \includegraphics{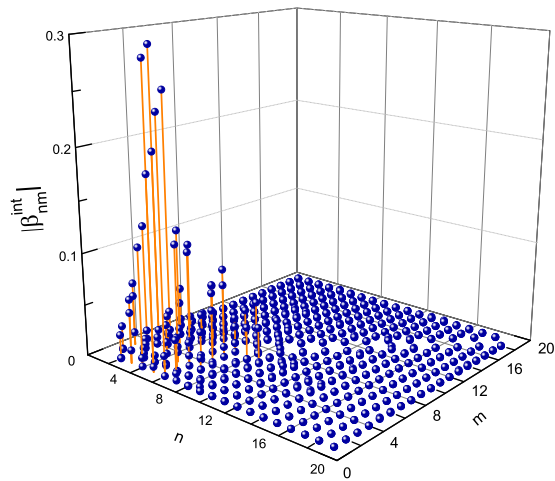}
  \caption{$\beta_{nm}^{int}$ for the energy class $j^{-2}$ when $\beta_{int} = 0.284$.}\label{fig:betanm}
\end{figure}
Interestingly, states $1$ and $3$ are responsible for $80\%$ of the hyperpolarizability.  Thus, while this outlier was not constrained to be a three-level system, three states dominate the response.  In contrast, when the hyperpolarizability is well below the limit, many states contribute.  The second entry in Table \ref{tab:betaSummary} shows that 8 excited states contribute to yield $\beta_{int} = 0.284$ for the $j^{-2}$ energy class for which $|f(E)| = 0.184$.  Figure \ref{fig:betanm} shows a plot of $\beta_{nm}^{int}$.  In this case, the dominant contribution is of opposite sign and of larger magnitude than of the total nonlinear response.  This illustrates how many states may contribute, but in such cases, the contributions can cancel.

\begin{figure}
  \includegraphics{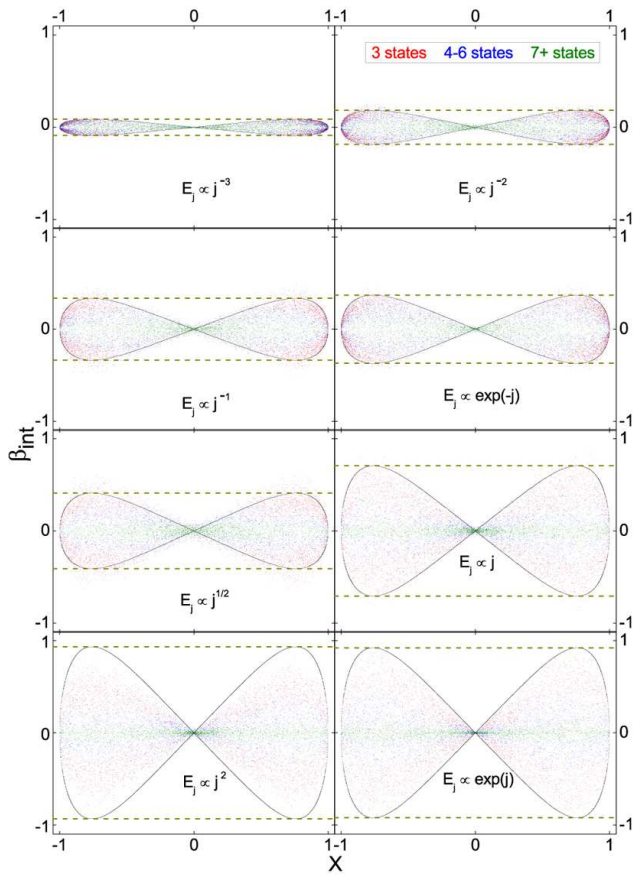}
  \caption{Density plots from 1 million Monte Carlo runs of $\beta_{int}$ as a function of $X$ for several energy classes of a 20-state model. The dashed lines represent $\pm f(E)$. Red, blue and green colors correspond to the $\beta_{int}$ values for which 3, 4 to 6 and 7+ states dominate the nonlinear response, respectively. The black solid curve represents a plot of $\pm f(E)G(X)$ for $E$ fixed by the energy class and as a function of X.}\label{fig:betaX}
\end{figure}
Figure \ref{fig:betaX} shows a density plot of the hyperpolarizability  as a function of $X(= x_{01}/x_{01}^{max})$ where $x_{01}^{max}$ is the fundamental limit of the transition moment defined in Equation \ref{x01max}.  For each energy class, the Monte Carlo simulations generate all possible values of $X$ in the $[-1,1]$ range as well as all the possible transition moments between all pairs of states.  Thus, the figure represents a sampling of the subset of the full Hilbert space for that energy class.  Associated with each energy class is a single value of $E$.  The plots in order of $j^{-3}$, $j^{-2}$, $j^{-1}$, etc. (left to right, top to bottom) are in order of decreasing $E$.

We can interpret this behavior by combining Equations \ref{betafG}-\ref{defG(X)}, which come from the three-level ansatz, to get
\begin{eqnarray}\label{betaint}
\beta_{int} &=& G(X)f(E)\nonumber \\
&=& \sqrt[4]{3}(1-E)^{3/2} \left( E^2 + \frac {3} {2} E + 1 \right)  X \sqrt{\frac {3} {2} \left( 1 - X^4\right)}.\nonumber \\
\end{eqnarray}
It is worthwhile to dwell on the origin of Equation \ref{betaint}.  It is derived from the three-level model of $\beta$ using the sum rules to reduce the number of parameters to two, namely $X$ and $E$.  While the hyperpolarizability is known to depend on the contributions from many excited states, it was proposed that when a quantum system is at the limit, only two excited states contribute.\cite{kuzyk10.01}  Thus, Equation \ref{betaint} applies, in principle, only at the limit.  The second assumption is that $X$ and $E$ are independent.  Under this condition, $f(E)$ and $G(X)$ are individually optimized and found to peak at unity for $E \rightarrow 0$ and $X = \sqrt[-4]{3} \approx 0.75$.

The solid black curve in Figure \ref{fig:betaX} represents $\pm f(E)G(X)$ for $E$ fixed by the energy class. Also shown is $\pm f(E)$ as horizontal dashed curves.  For each point on the plot, the number of dominant states is determined by the criteria that it contributes at least 25\% to the total, and color coded red for three states, blue for 4 to 6 states, and green for seven or more states.  First, we find that $f(E)$, aside from a small number of outliers, determines the range of $\beta_{int}$ values observed.  As $E$ gets smaller, more points in the distribution are closer to the limit.

The magnitude of the hyperpolarizability is also found to correlate with the number of dominant states.  For seven or more dominant state, the hyperpolarizability fills a narrow band centered on $\beta_{int} = 0$ and with a width of about 0.1.  The systems with four to six dominant states form a broader band while the three-level systems fill the range up to the values $f(E) G(X)$.  There are several interesting features of the data.  First, there is no reason for the function $f(E)$ to determine the upper bounds for systems with hyperpolarizabilities far from the limit.  However, it appears that the three-level results accurately predicts the upper bounds.  Thus, for a given energy class, the sum-rule constrained three-level model appears to yield the range of $\beta_{int}$ for each value of $X$.

All the points that fall on the curve $f(E) G(X)$ for every energy class represents a the three-level model.  Interestingly, this is the case not only for systems with an energy class for $\beta_{int} \approx 1$, but even when the energy function $f(E)$ limits the hyperpolarizability to 1/10 of the fundamental limit.  However, there are also three-state systems that are far below the $f(E) G(X)$ curve.  Thus, when a quantum system is described by three dominant states, this does not guarantee that the hyperpolarizability is at the fundamental limit even when the energy ratio and transition moments meet the criteria.  This is most likely due to many other excited states that are each contributing less than 25\% of the total, but in aggregate, decrease the value below the limit.

It is interesting that a plot calculated from the sum-rule-constrained three-level model so accurately provides a bound for all of the observed data.  In particular, in cases where the energy class yields a small value of $f(E)$, why are there no points outside the figure-eight pattern?  Furthermore, it is interesting that the hyperpolarizabilities all vanish when $X=0$.  While one may question if the three-level ansatz holds, our data here suggests that not only does it hold as originally posed, but the three-level ansatz is more broadly applicable than just at the limit.  Furthermore, the data suggests that indeed, $X$ and $E$ are independent, as previously postulated.

\subsection{Second Hyperpolarizability}
\begin{figure}
  \includegraphics{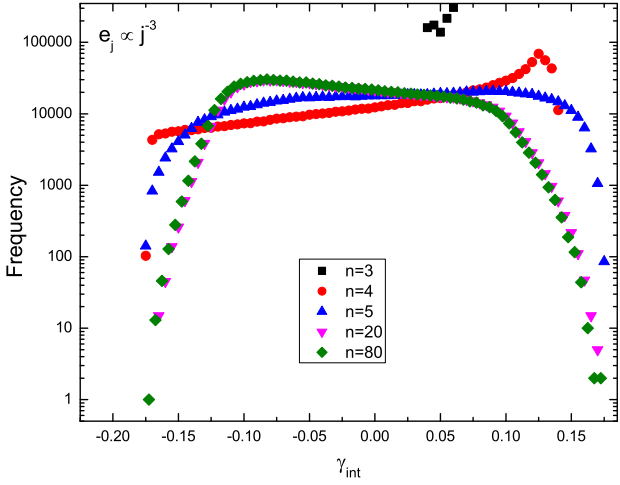}
  \caption{The distribution of the second hyperpolarizability for energy class $j^{-3}$.}\label{fig:Gamma,j^-3}
\end{figure}
In this section we focus on the second hyperpolarizability. Figure \ref{fig:Gamma,j^-3} shows the distributions of $\gamma_{int}$ for the energy class $j^{-3}$.  The range of the intrinsic second hyperpolarizability spans the range $-0.10 < \gamma_{int}< 0.15$.  However, the three-level model clusters around the value of $\gamma_{int} = 0.05$.  Thus, the three-level model gives a result that is far less than systems with more levels.  However, the four-level model peaks at the largest value of positive $\gamma_{int}$.  But, as we saw in the energy class $e^j$, shown in Figure \ref{Gamma,e^j}, the three-level model peaks at the limit for the optimized energy spectrum.

The three-level ansatz when applied to the off-resonant second order hyperpolarizability, $\gamma_{max}$ yields \cite{perez01.08},
\begin{equation}\label{gammaoff}
\gamma_{xxxx}^{off} = \frac{e^4 \hbar^4}{m^2 E_{10}^5}G_{\gamma}(E,X) ,
\end{equation}
where
\begin{eqnarray}\label{gGammaEX}
G_{\gamma}(E,X) &=& 4-5 \left(E-1\right)^2\left(E+1\right)\left(E^2+E+1\right)X^4 \nonumber \\
&&-2\left(E^2-1\right)E^3X^2-\left(E^3+E+3\right)E^2.\nonumber \\
\end{eqnarray}
The maximum value of $G_{\gamma}(E,X)$ is $4$ which is obtained for the extreme case $E=0$, leading to the fundamental limit of second hyperpolarizability, $\gamma_{max}$. The intrinsic value of any arbitrary system is found by dividing Equation \ref{gammaoff} by $\gamma_{max}$, defined in Equation \ref{gammamax},
\begin{equation}\label{gammaintgE}
\gamma_{int} = \frac{\gamma_{xxxx}^{off}}{\gamma_{max}} = \frac{G_\gamma(E,X)}{4} .
\end{equation}

\begin{figure}
  \includegraphics{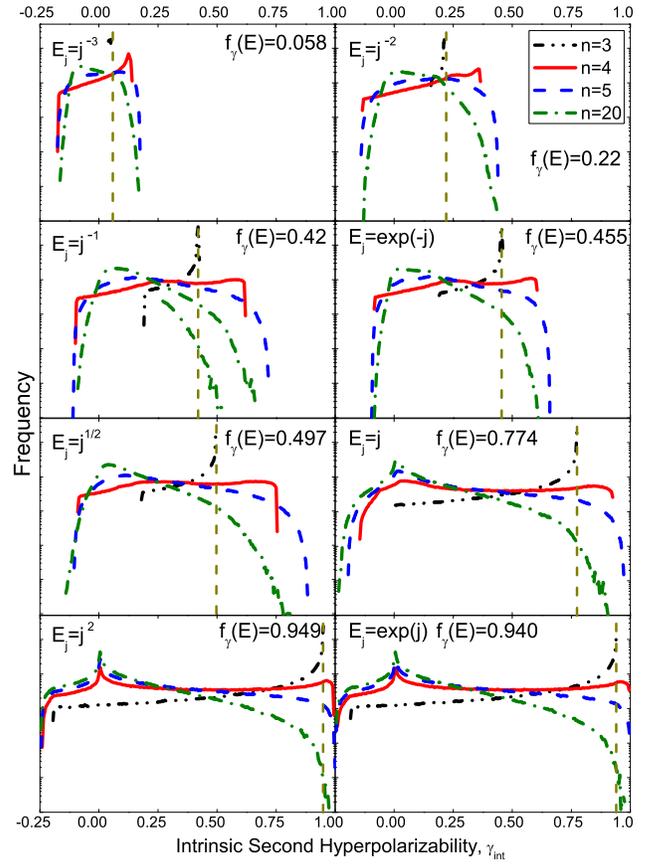}
  \caption{Distribution of $\gamma_{int}$ with energy for several different energy functions. The vertical dashed lines represent the energy function $f\gamma(E)$.}\label{fig:gammaSummary}
\end{figure}

In analogy to $\beta$, we are interested in the dependence of $G_{\gamma}(E,X)$ on $E$. Therefore we introduce $f(E) \equiv G(E,X_0(E))/4$ where $X_0(E)$ is the value of X that maximizes $G\left(E,X_0(E)\right)$ for fixed $E$. We find $X_0(E)$ by solving the equation $dG/dX=0$ for $X(E)$, yielding
\begin{equation}\label{X0E}
X_0^{(1)}(E) = 0 \quad \mbox{and} \quad X_0^{(2)}(E) = \pm \frac{E^{3/2}}{\sqrt{5\left(1 - E^3\right)}} .
\end{equation}
It is straightforward to show that the maximum value of $G_\gamma$ is obtained for $X_0^{(2)}(E)$. Introducing $X_0^{(2)}(E)$ in Equation \ref{gGammaEX} gives,
\begin{eqnarray}\label{fGammaE}
f_\gamma(E)&\equiv& \frac{G\left(E,X_0^{(2)}(E)\right)}{4}\nonumber \\
&=& \frac{1}{20} \left(19 - 14 E^2 - 6 E^3 - 4 E^5 + \frac{1 + E}{1 + E + E^2}\right) . \nonumber \\
\end{eqnarray}
For example, for the energy class $E_j=j^{-2}$ where $E=E_{10}/E_{20} = 0.843$, we find $f_\gamma(E) = 0.22$. The vertical dashed lines in Fig. \ref{fig:gammaSummary} represent $f_{\gamma}(E)$ values.

Figure \ref{fig:gammaSummary} summarizes the results of the second hyperpolarizability calculations for the various energy classes.  It is important to keep in mind that all the distributions for $\gamma_{int}$ are plotted on a log scale.  As the energy spectrum becomes more spread out, the range of $\gamma_{int}$ increases, where the range of $\gamma_{int}$ of the three-level model is smaller than the range in $\gamma_{int}$ with 4 or more states.  However, as the energy spectrum becomes more favorable, the three-state plot approaches the same range of $\gamma_{int}$ values as the other plots with a peak at $\gamma_{int}$ approaching unity.  In all simulations, the fundamental limit is not exceeded, suggesting that the three-level ansatz gives the correct limit.
\begin{table}\caption{Summary of the properties of the second hyperpolarizability for a sampling of energy functions when the second hyperpolarizability is small or near the limit.}\label{tab:gammaSummary}
\begin{tabular}{c c c c}
  \hline
  Function & $\gamma_{int}$ & Levels & Dominant States \\
  \hline
$j^{-2}$ & 0.443 & 5 & 1,2,3,4 \\
$j$ & 0.92 & 4 & 1,2,4 \\
$\exp(j)$ & 0.973 & 3 & 1,2 \\
\hline
\end{tabular}
\end{table}

Table \ref{tab:gammaSummary} gives a sampling of the largest $\gamma_{int}$ values, the dominant levels and their contribution to the intrinsic second hyperpolarizability for the three energy classes $E_j = \exp(j)$, $E_{j}=j$ and $E_{j} = j^{-2}$ for a $20$-state model. To find the contribution of pairs of states to $\gamma_{int}$, we use the missing state analysis \cite{Dirk89.01}. In this method, we calculate $\gamma_{int}$ in the absence of the pairs of states $i$ and $j$, which is called $\gamma_{ij}^{missing}$. The fractional difference between $\gamma_{ij}^{missing}$ and $\gamma_{int}$ describes the joint contribution of states $i$ and $j$ to $\gamma_{int}$. The smaller the value of $\gamma_{ij}^{missing}$, the larger the contribution of states $i$ and $j$ to the second hyperpolarizability.\cite{shafei10.01} For example, for $E_j = \exp(j)$, $\gamma_{1,2}^{missing}$ is on the order of $10^{-8}$ and $\gamma_{1,4}^{missing}$ is the second most significant term, which is of the order of $10^{-6}$ - indicating that the pair $(1,2)$ contributes to the second hyperpolarizability $100$ times more than the pair $(1,4)$.

For $E_j = j^{-2}$, 5 states contribute significantly to $\gamma$ with $(2,3)$ having the largest contribution. The energy class $E_{j} = j$ has a larger value of $E$ than $E_j = j^{-2}$, but only four states dominate. For $E_j = \exp(j)$, for which the second hyperpolarizability is closest to the limit of the three cases, only three states (ground state and two excited states) contribute to $\gamma$, as predicted by three level ansatz.

\begin{figure}
  \includegraphics{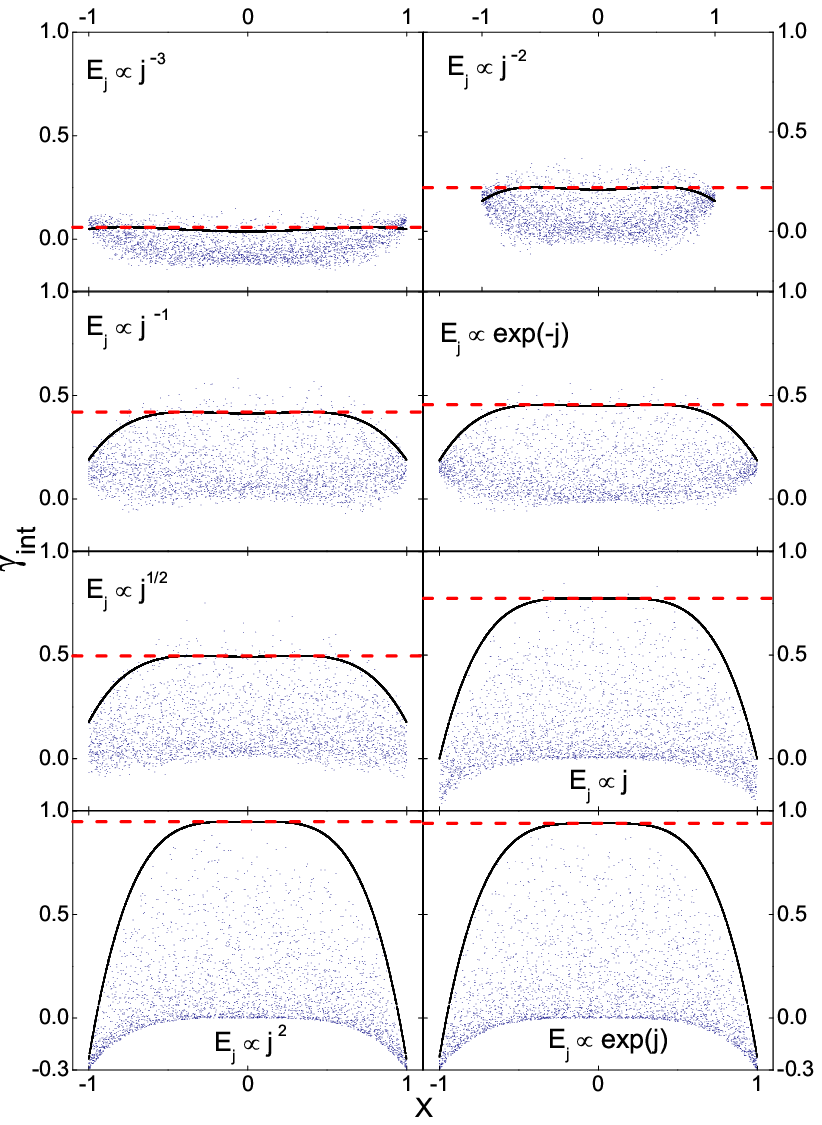}
  \caption{$\gamma_{int}$ as a function of $X$ for several energy classes using a 20-state model. Dashed lines represent $F_{\gamma}(E)$. The black solid curve specifies the limit defined by the sum-rule-constrained three-level model.}\label{fig:gammaX}
\end{figure}
Figure \ref{fig:gammaX} illustrates the distribution of $\gamma_{int}$ as a function of $X$ for a $20$-state model. While the Monte Carlo simulations generate all possible $X$ values, the resultant $\gamma_{int}$ range depends on the energy class. When $E$ is large, e.g. $E_j = j^{-3}$ and $E_j = j^{-2}$, the largest intrinsic second hyperpolarizabilities do not exceed $0.3$ while for classes such as $E_j=j^2$, they approach the limit. The dashed lines in Figure \ref{fig:gammaX} are the maximum attainable values of $\gamma_{int}$ in the sum-rule restricted three-level model, i.e. $f_{\gamma}(E)$. The numerical values of  $f_{\gamma}(E)$ are given in Figure \ref{fig:gammaSummary}.

When the system is not close to the fundamental limit, there are $\gamma_{int}$ values that exceed $f_{\gamma} (E)$ as represented by the vertical dashed lines in Figure \ref{fig:gammaSummary}, indicating that the sum-rule restricted three-level model does not necessarily produce the largest $\gamma_{int}$ values when the system is not optimized. Otherwise, when $E \rightarrow 0$, the three-level model leads to $\gamma_{int} \rightarrow 1$ as is the case for $E_j=e^{j}$ in Figure \ref{fig:gammaSummary}.

The curved solid lines in Figure \ref{fig:gammaX} plot $G_{\gamma}(X,E)/4$ for the value of $E$ that is defined for each energy class, indicating the largest $\gamma_{int}$ values that can be generated by each energy class when they are restricted to a sum-rule-constrained three-level model. When the three-level model is not optimized, i.e. $ G(X,E)\neq 4 $, the distribution of $\gamma_{int}$ exceeds $G_{\gamma}(X,E)/4$. However,  as the energy parameter $E$ gets smaller, $G_{\gamma}(X,E)/4$ increases and fewer points in the distribution fall outside the energy function.  When $E=0$, all points in the distribution fall below $G_{\gamma}(X,0)/4$, which defines the fundamental limit of the hyperpolarizability. This behavior supports the validity of three-level ansatz in calculating the fundamental limits of the second hyperpolarizability.

\subsection{The Gap}

The gap between the experimental results and the fundamental limit for the first\cite{kuzyk08.01} and second\cite{shafei10.01} order hyperpolarizabilities have been extensively discussed in the literature in which Monte Carlo simulations suggest that the gap might be due to the unfavorable arrangement of the excited state energies.

Figure \ref{Beta,j^-3} shows the distribution of the energy class $j^{-2}$ and $j^{-3}$, which approximately represents the energy spacing of atoms and molecules.  Most of the hyperpolarizabilities fall in the range $ -0.1 < \beta_{int} < +0.1$ for class $j^{-3}$ and $ -0.1 < \beta_{int} < +0.1$ for class $j^{-2}$. Thus, since most organic molecules fall in this range of energy class, one would expect that the very best molecules would fall below $\left| \beta_{int} \right| \approx 0.2$.
\begin{figure}
  \includegraphics{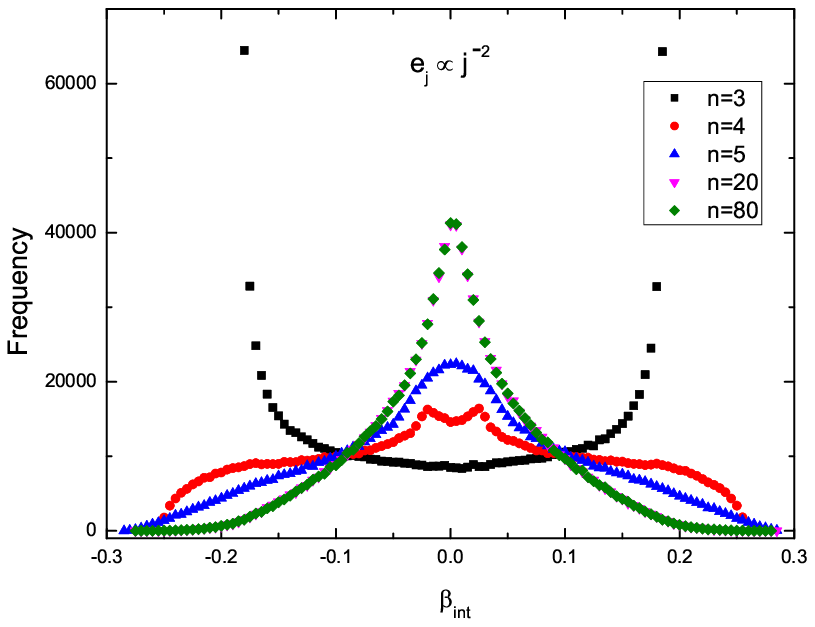}
  \includegraphics{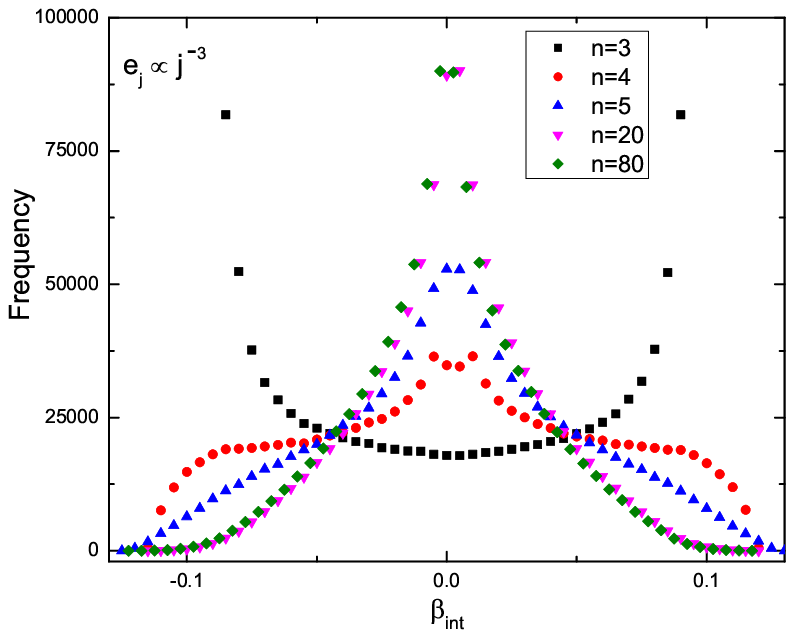}
  \caption{The distribution of $\beta_{int}$ for the energy class $j^{-2}$ (top) and $j^{-3}$ (bottom).}\label{Beta,j^-3}
\end{figure}

Assuming that the transition moments vary more between molecules than does the energy spacing,\cite{Tripa04.01,perez07.02} the statistically most likely observed hyperpolarizabilities are those near the peak in the distribution function.  The energy classes $j^{-2}$ and $j^{-3}$ both have a double-peaked structure in the range $ -0.03 < \beta_{int} < +0.03$ for the sum-rule constrained three-level model.  When more states contribute, the double-peaked behavior is not observed, but the width of the peak remains at 0.03.  Thus, based on statistics alone, one would expect most molecules to have an intrinsic hyperpolarizability of less than 0.03 - precisely the value of the observed gap.

At the heart of the gap may be the nature of coulomb forces, which lead to an energy spacing that is not conducive to optimizing the nonlinear-optical response unless the effective potential energy function is interrupted in a way to change the energy spacing of the lowest-energy eigenstates.

In the case of the second hyperpolarizability, the spread in $\gamma_{int}$ is larger than for $\beta_{int}$.  Furthermore, the distribution is flatter so it is statically more likely to find molecules with larger intrinsic second hyperpolarizability.  For the energy class $j^{-2}$, for example, the distribution narrows as the number of states in increased, but in all cases, the range is at least  $-0.07 <\gamma_{int} < 0.2$.  This is consistent with reports of $\gamma_{int} \approx 0.2$ by May and coworkers.\cite{May07.01}

Finally, in numerical simulations in which the potential energy function is varied to optimize $\beta_{int}$, the largest values observed are $\beta_{int} \approx 0.708$.\cite{zhou06.01,zhou07.02,zhou08.01,kuzyk10.01}  These simulations show that at the limit, potentials share certain universal properties, such as the energy ratio $E_{10} / E_{20} \approx 0.48$.  This corresponds to the energy class $j$, which shows a sharp drop-off beyond $\beta_{int} = 0.707$.  Thus, it appears that it may not be possible to reach the fundamental limit by varying the potential alone in  1D molecule.

\section{Conclusion}

Classifying Monte Carlo simulations using an energy spectrum function resolves several long-standing questions.  First, our work shows the centrality of energy spacing in determining the intrinsic nonlinear response.  While a broad range of transition moments are observed in atoms and molecules, the energy spacing - as characterized by the energy parameter, $E$, varies little between systems.  Indeed, the importance of the energy parameter in attaining larger hyperpolarizabilities has been demonstrated in several experimental studies.\cite{Tripa04.01,perez07.02}

The factor of 20-30 gap between the best molecules and the fundamental limit of the hyperpolarizability can be understood statistically.  When the energy spacing is characterized by $E=0$, then a random sampling of quantum systems shows an approximately flat distribution of hyperpolarizabilities from $\beta_{int} = 0$ to $\beta_{int} = 1$.  However, when the energy function is $E \approx 0.8$, as it is in typical molecules, a random sampling of transition moments would yield hyperpolarizabilities that fall a factor of 20 to 30 below the limit.  The largest possible hyperpolarizability in this case is about 0.1 to 0.2.  Unless quantum systems are identified that have a more favorable spacing, the intrinsic hyperpolarizability can be improved by no more than about a factor of 5.

The second-order hyperpolarizability, on the other hand, shows a similar behavior but with less sensitivity to the energy parameter.  For $E \approx 0.8$, $\gamma_{int}$ can be as large as about 0.2.  This is consistent with the identification of small molecules with intrinsic second hyperpolarizabilities in this range.\cite{May05.01,May07.01} Thus, when viewed in terms of energy classes, our Monte Carlo simulations explain the observed gap between the best molecules and the fundamental limit.

Studies aimed at finding the optimum potential energy functions that maximize the intrinsic hyperpolarizability find many potential functions that represent a local maximum of $\beta_{int}$.\cite{kuzyk09.01,kuzyk10.01}  Quantum systems that are so optimize all share certain universal properties, such as $E \approx 0.48$ and $\beta_{int} \approx 0.708$.  In our present studies, we find that the energy class $E_j = j$, with $E = 0.5$ is restricted to $\beta_{int} < 0.708$.  As such, our work suggests that solutions to the Schrodinger Equation with a potential energy function demands that the energy parameter $E > 0.48$.  Thus, in the process of optimizing the potential energy function, the energy parameter is optimized to its minimum possible value.  This is consistent with the suggestion that systems with more exotic Hamiltonians may be required to attain the fundamental limit.\cite{kuzyk08.01}

All of our simulations have shown the validity of the three-level ansatz - that is, at the fundamental limit, the system is represented by a sum-rule constrained three-level model with all of the oscillator strength shared by at most two states.  We find no instances where more than three states contribute at the limit, though a three-state model may describe a system that is far from the limits.  Also, since no values are found to exceed the limit, the theoretical foundations of the limit calculations appear to be on solid ground.

More interestingly, the sum-rule constrained three-level model appears to be more broadly applicable than just at the limits.  Even when the quantum system's hyperpolarizability is far from the limit, the energy function $f(E)$ defines the limits of the nonlinear response, as one would expect if the functions $f(E)$ and $G(X)$ -- which define the intrinsic hyperpolarizability according to $\beta_{int} = f(E) G(X)$ -- are independent.  Similarly, G(X) also provides a constraints, such as forcing $\beta_{int}$ to be small when $X$ is small.  In all cases, we find that as the number of states that contribute to the hyperpolarizability increases, the intrinsic hyperpolarizability is limited to narrower and narrower bands.   Similar behavior has been observed for the second hyperpolarizability.

Monte Carlo calculation using the energy classification scheme have bridged the divide between Monte Carlo simulations and potential energy optimization studies.  The power of the Monte-Carlo technique lies in the fact that all possible Hilbert spaces are probed, leading to very broad and fundamental relationships.  Using energy classifications allows the parameter space to be reduced to subsets that describe atoms and molecules.  Future refinements may lead to more specific design guidelines for making improved molecules for a variety of applications.  The potential for discovering new fundamental science with this approach is of equal importance.


\begin{thebibliography}{99}
\newcommand{\enquote}[1]{``#1''}

\bibitem{kuzyk00.01}
M.~G. Kuzyk, \enquote{{Physical Limits on Electronic Nonlinear Molecular
  Susceptibilities},} Phys. Rev. Lett. \textbf{85}, 1218 (2000).

\bibitem{kuzyk00.02}
M.~G. Kuzyk, \enquote{{Fundamental limits on third-order molecular
  susceptibilities},} Opt. Lett. \textbf{25}, 1183 (2000).

\bibitem{kuzyk01.01}
M.~G. Kuzyk, \enquote{{Quantum limits of the hyper-Rayleigh scattering
  susceptibility},} IEEE Journal on Selected Topics in Quantum Electronics
  \textbf{7}, 774 --780 (2001).

\bibitem{kuzyk03.01}
M.~G. Kuzyk, \enquote{{Fundamental limits on third-order molecular
  susceptibilities: erratum},} Opt. Lett. \textbf{28}, 135 (2003).

\bibitem{kuzyk03.02}
M.~G. Kuzyk, \enquote{{Erratum: Physical Limits on Electronic Nonlinear
  Molecular Susceptibilities},} Phys. Rev. Lett. \textbf{90}, 039902 (2003).

\bibitem{Tripa04.01}
K.~Tripathy, J.~P\'{e}rez~Moreno, M.~G. Kuzyk, B.~J. Coe, K.~Clays, and A.~M.
  Kelley, \enquote{{Why hyperpolarizabilities Fall Short of the Fundamental
  Quantum Limits},} J. Chem. Phys. \textbf{121}, 7932--7945 (2004).

\bibitem{Tripa07.01}
K.~Tripathy, J.~P\'{e}rez~Moreno, M.~G. Kuzyk, B.~J. Coe, K.~Clays, and A.~M.
  Kelley, \enquote{{Erratum: Why Hyperpolarizabilities Fall Short of the
  Fundamental Quantum Limits},} J . Chem. Phys. \textbf{125}, 079905 (2006).

\bibitem{slepk04.01}
A.~D. Slepkov, F.~A. Hegmann, S.~Eisler, E.~Elliot, and R.~R. Tykwinski,
  \enquote{{The surprising nonlinear optical properties of conjugated polyyne
  oligomers},} J. Chem. Phys. \textbf{120}, 6807--6810 (2004).

\bibitem{May05.01}
J.~C. May, J.~H. Lim, I.~Biaggio, N.~N.~P. Moonen, T.~Michinobu, and
  F.~Diederich, \enquote{{Highly efficient third-order optical nonlinearities
  in donor-substituted cyanoethynylethene molecules},} Opt. Lett. \textbf{30},
  3057--3059 (2005).

\bibitem{May07.01}
J.~C. May, I.~Biaggio, F.~Bures, and F.~Diederich, \enquote{{Extended
  conjugation and donor-acceptor substitution to improve the third-order
  optical nonlinearity of small molecules},} App. Phys. Lett. \textbf{90},
  251106 (2007).

\bibitem{wang04.01}
Q.~Y. Chen, L.~Kuang, Z.~Y. Wang, and E.~H. Sargent, \enquote{{Cross-linked
  C-60 polymer breaches the quantum gap},} Nano. Lett. \textbf{4}, 1673--1675
  (2004).

\bibitem{zhou08.01}
J.~Zhou and M.~G. Kuzyk, \enquote{{Intrinsic Hyperpolarizabilities as a Figure
  of Merit for Electro-optic Molecules},} J. Phys. Chem. C. \textbf{112},
  7978--7982 (2008).

\bibitem{kuzyk09.01}
M.~G. Kuzyk, \enquote{Using fundamental principles to understand and optimize
  nonlinear-optical materials,} J. Mat. Chem. \textbf{19}, 7444--7465 (2009).

\bibitem{Kang05.01}
H.~Kang, A.~Facchetti, P.~Zhu, H.~Jiang, Y.~Yang, E.~Cariati, S.~Righetto,
  R.~Ugo, C.~Zuccaccia, A.~Macchioni, C.~L. Stern, Z.~Liu, S.~T. Ho, and T.~J.
  Marks, \enquote{{Exceptional Molecular Hyperpolarizabilities in Twisted
  $\pi$-Electron System Chromophores},} Angew. Chem. Int. Ed. \textbf{44},
  7922--7925 (2005).

\bibitem{Kang07.01}
H.~Kang, A.~Facchetti, H.~Jiang, E.~Cariati, S.~Righetto, R.~Ugo, C.~Zuccaccia,
  A.~Macchioni, C.~L. Stern, Z.~F. Liu, S.~T. Ho, E.~C. Brown, M.~A. Ratner,
  and T.~J. Marks, \enquote{{Ultralarge hyperpolarizability twisted
  $\pi$-electron system electro-optic chromophores: Synthesis, solid-state and
  solution-phase structural characteristics, electronic structures, linear and
  nonlinear optical properties, and computational studies},} J. Am. Chem. Soc.
  \textbf{129}, 3267--3286 (2007).

\bibitem{kuzyk10.01}
M.~G. Kuzyk, \enquote{A bird's-eye view of nonlinear-optical processes:
  Unification through scale invariance,} Nonliner Optics Quantum Optics
  \textbf{40}, 1--13 (2010).

\bibitem{rober09.01}
R.~Roberts, T.~Schwich, T.~Corkery, M.~Cifuentes, K.~Green, J.~Farmer, P.~Low,
  T.~Marder, M.~Samoc, and M.~Humphrey, \enquote{Organometallic complexes for
  nonlinear optics. 45. dispersion of the third-order nonlinear optical
  properties of triphenylamine-cored alkynylruthenium dendrimers,} Advanced
  Materials \textbf{21}, 2318--2322 (2009).

\bibitem{kuzyk03.03}
M.~G. Kuzyk, \enquote{{Fundamental Limits on Two-Photon Absorption
  Cross-Sections},} J. Chem Phys. \textbf{119}, 8327--8334 (2003).

\bibitem{perez05.01}
J.~P\'{e}rez~Moreno and M.~G. Kuzyk, \enquote{{Fundamental limits of the
  dispersion of the two-photon absorption cross section},} J. Chem. Phys.
  \textbf{123}, 194101 (2005).

\bibitem{perez11.01}
J.~P\'{e}rez-Moreno and M.~G. Kuzyk, \enquote{A correspondence on
  "organometallic complexes for nonlinear optics. 45. dispersion of the
  third-order nonlinear optical properties of triphenylamine-cored
  alkynylruthenium dendrimers". increasing the nonlinear optical response by
  two orders of magnitude.} Advanced M \textbf{DOI: 10.1002/adma.201003421}
  (2011).

\bibitem{kuzyk06.02}
M.~G. Kuzyk and D.~S. Watkins, \enquote{{The Effects of Geometry on the
  Hyperpolarizability},} J. Chem Phys. \textbf{124}, 244104 (2006).

\bibitem{zhou06.01}
J.~Zhou, M.~G. Kuzyk, and D.~S. Watkins, \enquote{{Pushing the
  hyperpolarizability to the limit},} Opt. Lett. \textbf{31}, 2891 (2006).

\bibitem{zhou07.02}
J.~Zhou, U.~B. Szafruga, D.~S. Watkins, and M.~G. Kuzyk, \enquote{{Optimizing
  potential energy functions for maximal intrinsic hyperpolarizability},} Phys.
  Rev. A \textbf{76}, 053831 (2007).

\bibitem{perez07.01}
J.~P\'{e}rez-Moreno, Y.~Zhao, K.~Clays, and M.~G. Kuzyk, \enquote{{Modulated
  conjugation as a means for attaining a record high intrinsic
  hyperpolarizability},} Opt. Lett. \textbf{32}, 59--61 (2007).

\bibitem{perez09.01}
J.~P\'{e}rez-Moreno, Y.~Zhao, K.~Clays, M.~G. Kuzyk, Y.~Shen, L.~Qiu, J.~Hao,
  and K.~Guo, \enquote{Modulated conjugation as a means of improving the
  intrinsic hyperpolarizability,} J. Am. Chem. Soc. \textbf{131}, 5084–5093
  (2009).

\bibitem{watkins09.01}
D.~S. Watkins and M.~G. Kuzyk, \enquote{Optimizing the hyperpolarizability
  tensor using external electromagnetic fields and nuclear placement,} J. Chem.
  Phys. \textbf{131}, 064110 (2009).

\bibitem{wangm06.01}
M.~Wang, X.~Hu, D.~N. Beratan, and W.~Yang, \enquote{{Designing molecules by
  optimizing potentials},} J. Am. Chem. Soc. \textbf{128}, 3228--3232 (2006).

\bibitem{kuzyk08.01}
M.~C. Kuzyk and M.~G. Kuzyk, \enquote{{Monte Carlo Studies of the Fundamental
  Limits of the Intrinsic Hyperpolarizability},} J. Opt. Soc. Am. B.
  \textbf{25}, 103--110 (2008).

\bibitem{shafei10.01}
S.~Shafei, M.~C. Kuzyk, and M.~G. Kuzyk, \enquote{Monte carlo studies of the
  intrinsic second hyperpolarizability,} J. Opt. Soc Am. B \textbf{27},
  1849--1856 (2010).

\bibitem{hefli98.01}
J.~R. Heflin, K.~Y. Wong, O.~Zamani-Khamiri, and A.~F. Garito,
  \enquote{{Symmetry-Controlled Electron Correlation Mechanism for Third Order
  Nonlinear Optical Properties of Conjugated Linear Chains},} Mol. Cryst. Liq.
  Cryst. \textbf{160}, 37 (1988).

\bibitem{hefli98.02}
J.~R. Heflin, K.~Y. Wong, O.~Zamani-Khamiri, and A.~F. Garito,
  \enquote{{Nonlinear optical properties of linear chains and
  electron-correlation effects},} Phys. Rev. B \textbf{38}, 1573–1576 (1988).

\bibitem{wu89.01}
J.~W. Wu, J.~R. Heflin, R.~A. Norwood, K.~Y. Wong, O.~Zamani-Khamiri, A.~F.
  Garito, P.~Kalyanaraman, and J.~Sounik, \enquote{{Nonlinear-optical Processes
  in Lower-dimensional Conjugated Structures},} J. Opt. Soc. Am. B \textbf{6},
  707--20 (1989).

\bibitem{Goldman82.01}
S.~P. Goldman and G.~W.~F. Drake, \enquote{Relativistic sum rules and integral
  properties of the dirac equation,} Phys. Rev. A \textbf{25} (1982).

\bibitem{Leung86.01}
P.~T. Leung and M.~L. Rustgi, \enquote{Relativistic corrections to bethe sum
  rule,} Phys. Rev. A \textbf{33} (1986).

\bibitem{cohen04.01}
S.~M. Cohen, \enquote{Aspects of relativistic sum rules,} Advances in Quantum
  Chemistry \textbf{46}, 241--265 (2004).

\bibitem{keina08.01}
S.~Keinan, M.~J. Therien, D.~N. Beratan, and W.~T. Yang, \enquote{{Molecular
  Design of Porphyrin-Based Nonlinear Optical Materials},} J. Phys. Chem. A
  \textbf{112}, 12203--12207 (2008).

\bibitem{perez07.02}
J.~P\'{e}rez-Moreno, I.~Asselberghs, Y.~Zhao, K.~Song, H.~Nakanishi, S.~Okada,
  K.~Nogi, O.-K. Kim, J.~Je, J.~Matrai, M.~De~Mayer, and M.~G. Kuzyk,
  \enquote{{Combined molecular and supramolecular bottom-up nano-engineering
  for enhanced nonlinear optical response: Experiments, modelling and
  approaching the fundamental limit},} J. Chem. Phys. \textbf{126}, 074705
  (2007).

\bibitem{kuzyk05.02}
M.~G. Kuzyk, \enquote{{Compact sum-over-states expression without dipolar terms
  for calculating nonlinear susceptibilities},} Phys. Rev. A \textbf{72},
  053819 (2005).

\bibitem{perez01.08}
J.~P\'{e}rez-Moreno, K.~Clays, and M.~G. Kuzyk, \enquote{{A new dipole-free
  sum-over-states expression for the second hyperpolarizability},} J. Chem.
  Phys. \textbf{128}, 084109 (2008).

\bibitem{Dirk89.01}
C.~W. Dirk and M.~G. Kuzyk, \enquote{{Missing-state analysis: A method for
  determining the origin of molecular nonlinear optical properties},} Phys.
  Rev. A \textbf{39}, 1219--1226 (1989).
\end{thebibliography}

\end{document}